\documentclass{IOS-Book-Article}

\normalfont
\usepackage[T1]{fontenc}
\usepackage{graphicx}
\begin{document}
\begin{frontmatter}                           % The preamble begins here.

%\pretitle{Pretitle}
\title{Linear-optics manipulations\\
of photon-loss codes}
\runningtitle{Linear-optics manipulations}
%\subtitle{Subtitle}

\author{\fnms{Konrad} \snm{Banaszek}%
\thanks{Corresponding author: ul. Grudzi\k{a}dzka 5, PL-87-100 Toru\'{n},
Poland; E-mail: kbanasz@fizyka.umk.pl.}},
\author{\fnms{Wojciech} \snm{Wasilewski}}

\runningauthor{K. Banaszek et al.}
\address{Institute of Physics, Nicolaus Copernicus University, Toru\'{n}, Poland}

\begin{abstract}
We discuss codes for protecting logical qubits carried by
optical fields from the effects of amplitude damping, i.e.\ linear photon loss.
We demonstrate that the correctability condition for one-photon loss imposes limitations on the range of manipulations
than can be implemented with passive linear-optics networks.
\end{abstract}

\end{frontmatter}

\thispagestyle{empty}
\pagestyle{empty}

\section{Introduction}
Quantum states encoded in optical fields are an obvious way to implement quantum communication protocols \cite{QComm}. The optical
approach offers also a route towards scalable quantum computing, with the particularly promising linear-optics scheme of Knill, Laflamme, and Milburn \cite{KLM}. Optical
fields are distinguished from standard qubit models of quantum information processing in two ways. First, photons are bosons that can occupy field modes in arbitrary numbers. Although a qubit can be implemented as a superposition of a single photon in two orthogonal modes, the entire Hilbert space describing optical fields has plenty of room to go beyond this standard, dual-rail representation. Secondly, most important error mechanisms that affect optical fields have a specific form. This enables one to optimize strategies for shielding quantum information from their deleterious effects.

The above features can be illustrated with the example of photon loss, also referred to in literature as amplitude damping. Such a mechanism can be modeled as a transmission of optical fields through a partly reflecting beam splitter. This attenuates the transmitted field, leading to a random removal of photons from the initial state. The effects of photon loss can be dealt with by adopting the strategy of quantum error correction \cite{QEC}, which consists in designing so-called code subspaces in which qubits are protected from dominant errors. Such codes can be constructed from states that contain more than one photon per mode \cite{CLY}, which makes them more efficient in terms of required numbers of photons and modes.

A natural method to manipulate optical codes is to use linear optics networks \cite{LinOpt}.
In the simplest scenario, such networks are passive, i.e.\ they do not involve auxiliary photons, conditional detection, or feed-forward operations. In this contribution we review the recent proof \cite{WB} that even the simplest codes, which correct for just a single photon loss, cannot be universally processed using passive linear optics only. As we will see below, the same properties that make the codes correctable for photon loss, also prohibit a range of linear-optics manipulations.

\section{Photon-loss codes}
\label{Sec:Photonlosscodes}

Logical qubits can be protected from the effects of errors by preparing more complex states of physical systems \cite{QEC}. Such states span the so-called code subspace, within which an arbitrary quantum superposition remains preserved despite the occurrence of an error from a certain class.
In the most elementary case, the subspace is spanned by two orthogonal states which we shall denote by $|L\rangle$ and $|H\rangle$. The specific scenario we shall consider here is shielding logical qubits carried by optical fields from amplitude damping by encoding them in suitable multiphoton states.
Let us restrict our attention to errors induced by the loss at most one photon from the field. If we make two assumptions:
\begin{itemize}
\item the code is constructed in a subspace with a fixed total photon number,
\item the damping parameter is identical for all the modes involved,
\end{itemize}
then the necessary and sufficient conditions for a code constructed in a system of bosonic modes to be robust against one-photon loss are given by \cite{CLY}:
\begin{equation}\label{Kn-La_conditions}
\langle H|\hat a_i^\dagger \hat a_j|L \rangle = 0, \qquad
\langle H|\hat a_i^\dagger \hat a_j|H \rangle = \langle L|\hat a_i^\dagger \hat a_j|L \rangle
\end{equation}
where $\hat{a}_i$ and $\hat{a}_i^\dagger$ denote respectively the annihilation and creation operators of the field modes, and the indices $i,j=1,\ldots, N$ run over all of $N$ bosonic modes. The action of an operator $\hat{a}_j$ on
the code states can be interpreted as an event when a third party has observed in the leaked portion of the field one photon in the $j$th mode. The third party belongs to the external environment, and her actions, as well as the outcome of her measurement are of course unknown to the owner of the qubit.

The equation $\langle H|\hat a_i^\dagger \hat a_j|L \rangle = 0$
implies a simple constraint: it must not be possible to pass between two orthogonal code states by moving just one photon between the modes,
i.e.\ annihilating it from the mode $\hat{a}_j$ and creating one in the mode $\hat{a}_i$. Starting from this observation, it is easy to construct two elementary examples of photon-loss codes
using four photons distributed between two modes \cite{CLY}:
\begin{equation}
\label{Eq:FourPhoton}
|L\rangle = \frac{1}{\sqrt{2}}(|04\rangle + |40\rangle), \qquad
|H\rangle = | 22 \rangle,
\end{equation}
and using three photons distributed between three modes \cite{WB}:
\begin{equation}
\label{Eq:ThreePhoton}
|L\rangle = \frac{1}{\sqrt{3}}(|003\rangle + |030\rangle + | 300 \rangle), \qquad
|H\rangle = | 111 \rangle.
\end{equation}
It is straightforward to verify that the correctability conditions are satisfied, and furthermore
that all the states $\hat{a}_1 |L\rangle, \hat{a}_1 |H\rangle, \hat{a}_2 |L\rangle, \hat{a}_2 |H\rangle,
\ldots$ are mutually orthogonal and have identical norms.

A code subspace can be conveniently characterized with the help of the corresponding projection operator $\hat{P}_{\cal C}$, which in our two-dimensional case takes the form $\hat{P}_{\cal C} = |L\rangle \langle L | +
|H\rangle \langle H|$. Then the correctability conditions can be written in a very compact form as \cite{Kribs}:
\begin{equation}
\label{Eq:CorrectPC}
\hat{P}_{\cal C} \hat{a}^\dagger_i \hat{a}_j \hat{P}_{\cal C} = G_{ij} \hat{P}_{\cal C},
\end{equation}
where $G=(G_{ij})$ is a certain $N\times N$ matrix. This matrix is hermitian, which is easily proven by considering
the hermitian conjugation of Eq.~(\ref{Eq:CorrectPC}).

\section{Linear optics transformations}

A general passive linear optics transformation of a system of $N$ modes can be written as:
\begin{equation}
\label{Eq:linopttransf}
\hat{a}_i = \sum_{k=1}^{N} \Gamma_{ik} \hat{b}_k
\end{equation}
where the operators $\hat{a}_i$ describe the input representation,
the operators $\hat{b}_j$ refer to the output representation, and $\Gamma = (\Gamma_{ij})$
is a unitary $N\times N$ matrix. The unitarity of the matrix $\Gamma$ guarantees the preservation of commutation
relations for the field operators. Any linear-optics transformation can be reversed, therefore the labels
of input and output representations are purely conventional.

Let us now write the correctability condition given in Eq.~(\ref{Eq:CorrectPC}) in the representation
of the output modes $\hat{b}_k$. A straightforward calculation shows that
\begin{equation}
\hat{P}_{\cal C} \hat{b}^\dagger_k \hat{b}_l \hat{P}_{\cal C} =
(\Gamma^T G \Gamma^\ast)_{kl} \hat{P}_{\cal C},
\end{equation}
where $\Gamma^T$ and $\Gamma^\ast$ denote respectively the
transposition and the complex conjugation of the matrix $\Gamma$.
This means that after the application of the network $\Gamma$
the subspace $\hat{P}_{\cal C}$ remains a photon-loss code, and that
the only change is the transformation of the matrix $G$ on the right hand side of the correctability condition.
This property reflects the fact that in our error model
the correctability condition is independent of the specific modal
decomposition. Indeed, the third party monitoring the leaked field can decompose it in an arbitrary basis
of modes and measure them individually for the presence of a photon. Because the damping coefficients are assumed
to be identical for all the modes, such a procedure performed on the leaked field does not alter the error model.

The fact that the photon-loss code is preserved by passive linear-optics transformations has important implications
for encoding and decoding. Suppose that we start from a qubit in the standard, dual-rail representation, with the aim
of mapping it onto the encoded subspace. A simple way to accomplish this would be to combine it
with auxiliary modes prepared in a certain state and apply a passive linear optics transformation.
However, the reversed version of the argument presented above implies
that if the output is a photon-loss code, then the input needs to be such a code as well. This means that the
input qubit itself is protected against photon loss, which obviously is not the case for the dual-rail representation.
The same applies to decoding: if we know a priori that no photon loss occurred, we cannot convert the encoded
qubit back into the dual-rail representation using a passive network. Therefore, there cannot exist a passive network that would work more universally for the input affected by errors and provide a decoded qubit with the error syndrome contained in the state of auxiliary modes.

\section{Single-qubit gates}
We will now consider single-qubit gates operating on the encoded qubit that can be implemented with passive networks.
Thus we are looking for networks that do not mix the code subspace $\hat{P}_{\cal C}$ with the remaining complement of the Hilbert space. Let us consider a linear-optics transformation of the annihilation operators given
by Eq.~(\ref{Eq:linopttransf}). The transformation of the modes induces a certain
unitary operator $\hat{R}(\Gamma)$ in the Hilbert space of the multimode bosonic system. The condition
that the operator $\hat{R}(\Gamma)$ does not take us beyond the code subspace can be written as:
\begin{equation}
\label{Eq:PRGammaP=RGammaP}
 \hat{R}(\Gamma) \hat{P}_{\cal C} = \hat{P}_{\cal C} \hat{R}(\Gamma) \hat{P}_{\cal C}.
\end{equation}
The set of all networks that preserve the code subspace forms a group. Let us now suppose that this group is continuous. This means that we can find a one-parameter subgroup composed of elements $\Gamma_s$
parameterized with a real parameter $s$ according to:
\begin{equation}
\label{Eq:Gammas=exp-sLambda}
\Gamma_s = \exp(-i s \Lambda)
\end{equation}
where $\Lambda = (\Lambda_{ij})$ is an $N\times N$ hermitian matrix. Mathematically, $\Lambda$ is an element of the Lie algebra associated with the Lie group of unitary $N\times N$ matrices.

The unitary operator $\hat{R}(\Gamma_s)$ can then be written as:
\begin{equation}
\hat{R}(\Gamma_s) = \exp( -i s \hat{R}(\Lambda))
\end{equation}
where $\hat{R}(\Lambda)$ is the representation of the matrix $\Lambda$ for the multimode bosonic system. It is given by a bilinear combination of the creation and annihilation operators \cite{Gilmore}:
\begin{equation}
\label{Eq:RLambda}
\hat{R}(\Lambda) = \sum_{i,j=1}^{N} \Lambda_{ij} \hat{a}_i^\dagger \hat{a}_j.
\end{equation}
The correctness of this expression can be verified by considering operators
$\hat{a}_i(s) = \exp(is\hat{R}(\Lambda))\hat{a}_i \exp(-is\hat{R}(\Lambda))$
and writing differential equations for $d\hat{a}_i(s)/ds$, whose solution recovers
Eq.~(\ref{Eq:linopttransf}) with the transformation matrix given by Eq.~(\ref{Eq:Gammas=exp-sLambda}).

Let us now consider an infinitesimal transformation of the form $\hat{R}(\Gamma_s) = \hat{I} - i s \hat{R}(\Lambda)$.
The second term, given explicitly in Eq.~(\ref{Eq:RLambda}), comprises a sum of expressions of the form $\hat{a}_i^\dagger \hat{a}_j$ that appear also in the correctability condition given in Eq.~(\ref{Eq:CorrectPC}).
It is easy to see that:
\begin{equation}
\hat{P}_{\cal C}\hat{R}(\Lambda)\hat{P}_{\cal C} = \sum_{i,j=1}^{N} \Lambda_{ij} \hat{P}_{\cal C}
\hat{a}_i^\dagger \hat{a}_j \hat{P}_{\cal C} =  \sum_{i,j=1}^{N} \Lambda_{ij}G_{ij} \hat{P}_{\cal C} =
\lambda \hat{P}_{\cal C},
\end{equation}
where we introduced a real coefficient $\lambda = \mbox{Tr}(\Lambda G^T)$. Inserting this result to Eq.~(\ref{Eq:PRGammaP=RGammaP}) yields:
\begin{equation}
\hat{R}(\Gamma_s)\hat{P}_{\cal C} = \hat{P}_{\cal C}\hat{R}(\Gamma_s)\hat{P}_{\cal C}
= \hat{P}_{\cal C} - i s \hat{P}_{\cal C}\hat{R}(\Lambda)\hat{P}_{\cal C} = (1-is\lambda) \hat{P}_{\cal C}
\end{equation}
This means that for general, not necessarily infinitesimal, operators $\hat{R}(\Gamma_s)$ we have
$\hat{R}(\Gamma_s)\hat{P}_{\cal C} = e^{-i s \lambda} \hat{P}_{\cal C}$.
Therefore the operator $\hat{R}(\Gamma_s)$ restricted to the code subspace generates only an irrelevant,
uniform phase factor.
Consequently, there does not exist a continuous group of linear transformations that would produce non-trivial gates
on the encoded qubit.

\section{Examples}

We demonstrated in the preceding section that groups of transformations which can be implemented on photon-loss codes using passive networks must be discrete. For two examples of codes presented in Section~\ref{Sec:Photonlosscodes} these groups can be found analytically. In the case of the four-photon code defined in Eq.~(\ref{Eq:FourPhoton}), one can use the parameterization of special unitary $2\times 2$ matrices in terms of Euler angles to show easily that all gates preserving the code subspace are generated by two transformations given, up to overall phase factors, by:
\begin{equation}
\Gamma_2 = \frac{1}{\sqrt{2}}\left(\begin{array}{cc}
1 & 1 \\
-1 & 1
\end{array}\right),
\qquad
\Gamma_2' = \left(\begin{array}{cc}
1 & 0 \\
0 & i
\end{array}\right).
\label{Eq:4phtr}
\end{equation}
The above transformations are realized respectively by a balanced beam splitter and a $\pi/2$ phase shift.
In order to gain an insight into the structure of the set of gates, it is helpful to consider transformations of the state $|H\rangle = |22\rangle$. The application of $\Gamma_2$ and $\Gamma_2'$ generates two other states given, up to phase factors, by $(\sqrt{3} |L\rangle - |H\rangle)/2$ and $(\sqrt{3} |L\rangle + |H\rangle)/2$, which together form an equilateral triangle in the Bloch sphere of the encoded qubit, shown in Figure~1(a). Gates that can be implemented with passive linear optics form the rotation group of this triangle.

A more lengthy, but still elementary reasoning \cite{WB} shows that for the three-photon code defined in Eq.~(\ref{Eq:ThreePhoton}) the transformations preserving the code subspace are obtained from two generators:
\begin{equation}
\Gamma_2 =\frac{1}{\sqrt{3}}\left(\begin{array}{ccc}
1 & 1 & 1 \\
1 & e^{2\pi i/3} & e^{-2\pi i/3} \\
1 & e^{-2\pi i/3} & e^{2\pi i/3} \end{array}\right), \qquad \Gamma_3 = \left(\begin{array}{ccc}
1 & 0 & 0 \\
0 & 1 & 0 \\
0 & 0 & e^{2\pi i/3} \end{array}\right)
\label{Eq:3phtr}
\end{equation}
which correspond respectively to a tritter \cite{Tritter} and a $2\pi/3$ phase shift on one of the modes.
Starting from the initial state $|H\rangle = |111\rangle$, these transformations generate a regular tetrahedron in the Bloch sphere of the encoded qubit, with vertices corresponding to the states $|H\rangle, (\sqrt{2}|L \rangle - |H\rangle )/\sqrt{3}$, and
$(\sqrt{2}|L \rangle - e^{\pm 2 \pi i/3}|H\rangle)/\sqrt{3}$, shown in Figure~1(b). As before, passive networks realize the rotation group of this solid.

\begin{figure}[t]
%\leavevmode\fbox{\parbox[b][30mm][s]{30mm}{
%\vfill\footnotesize Please include graphics in Encapsulated
%PostScript (.eps) format\vfill}}
\center\includegraphics[scale=0.2]{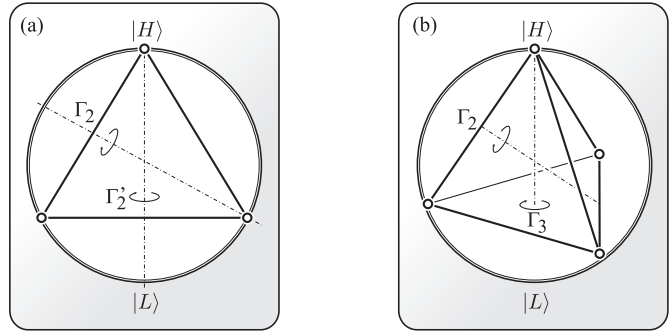} \caption{The Bloch spheres of the logical qubit for (a) the
four-photon code defined in Eq.~(\ref{Eq:FourPhoton}) and (b) the three photon code defined in
Eq.~(\ref{Eq:ThreePhoton}). The points represent states that can be produced from the logic state $|H\rangle$
using passive networks, and dashed lines depict rotations that are generated by unitary transformations
defined respectively in Eqs.~(\ref{Eq:4phtr}) and (\ref{Eq:3phtr}).}
\end{figure}

\section{Conclusions}
We have shown that restrictions on manipulating photon-loss codes with linear optics
are intimately linked to the correctability condition itself. The invariance of the correctability condition with respect to unitary transformations realized by
passive networks prohibits their use for encoding and decoding. Furthermore,
passive linear-optics networks are obtained from infinitesimal generators of the form $\hat{a}_i^\dagger \hat a_{j}$ that either move one photon between modes, or introduce linear phase shifts. However, in the code subspace these expressions need to reduce to $c$-numbers to ensure correctability, which severely limits available manipulations on encoded qubits.

\section*{Acknowledgements}
This work has been supported by the European Commission under the Integrated Project Qubit
Applications (QAP) funded by the IST directorate as Contract Number 015848, Polish MNiSW grant
1~P03B~011~29 and AFOSR under grant number FA8655-06-1-3062.

\end{document}